\documentstyle[aps,prl,preprint]{revtex}
\def\d{{\rm d}}  
\begin{document}

\renewcommand{\theparagraph}{\Alph{paragraph}}

\tightenlines 

\title{A Mechanical Model of Normal and Anomalous Diffusion \\}

\author{H. Kunz $^{(1)}$, R. Livi$^{(2)}$, 
         and A. S\"ut\H o$^{(3)}$\\[-2mm]$ $}

\address{
        $^{1}$
        Institut de Physique Theorique, Ecole Polytechnique Federale de
        Lausanne\\ CH-1015 Lausanne, Switzerland\\
        $^{2}$
        Dipartimento di Fisica, Universit\'a di Firenze, 
        and INFM, UdR Firenze\\ via Sansone 1  I-50019 Firenze, Italy\\
        $^{3}$
        Research Institute for Solid State Physics and Optics of the
        Hungarian Academy of Sciences\\ P. O. Box 49, H-1525
        Budapest 114, Hungary\\
        }

\date{Draft version, \today}

\maketitle

\begin{abstract}
The overdamped dynamics of a charged particle driven by an
uniform electric field through a random sequence of scatterers in
one dimension is investigated. Analytic expressions of the mean
velocity and of the velocity power spectrum are presented.
These show that above a threshold value of the field
normal diffusion is superimposed to ballistic motion.
The diffusion constant can be given explicitly.
At the threshold field
the transition between conduction and localization
is accompanied by an anomalous diffusion. Our results
exemplify that, even in the absence of time-dependent stochastic forces,
a purely mechanical model equipped with a
quenched disorder can exhibit normal as well as anomalous
diffusion, the latter emerging as a critical property.
\vskip .2 cm
\noindent
{PACS number(s): 05.40.-a}
\end{abstract}

\newpage

\section{Introduction}

Various dynamical models have been introduced for describing
anomalous diffusion. A good deal of them rely upon a
stochastic formulation of the dynamical rule, as in the celebrated
paper by Ya. G. Sinai, describing a random walk in a 1d random
potential \cite{Sinai}. More generally, the dynamics is either given by a 
master equation, with random hopping rates, describing the effect
of a random environment, or, in the continuous version, by a 
Langevin equation with a time--dependent random force, usually
assumed as Gaussian. The literature in this field is so huge that here we limit
ourselves to mention the excellent review article by Bouchaud
and Georges \cite{BG}. 

Anomalous as well as normal diffusion have been tackled
also in the framework of deterministic dynamics. A well-known
example is given by the standard map \cite{Ott,Lichten}, where normal
diffusion is observed when the phase space is represented by a 
chaotic sea with sparse stability island \cite{Lichten,Ishizaki}. 
Conversely, when many stability islands coexist finite time
trapping of chaotic orbits around them induces correlations yielding
anomalous diffusion \cite{Ishizaki}. Strong anomalous diffusion has been 
observed in other deterministic systems: here we just mention two models
of 1d intermittent maps introduced by Geisel et al. \cite{Geisel}
and by Pikovsky \cite{Pikov}. In the former example the complex
scenario of normal and anomalous diffusion emerging from a 
chaotic dynamics was pointed out for the first time.
In the latter example, it has been shown
that one can pass from normal to anomalous diffusion while varying the 
polynomial behaviour at the unstable fixed point of the intermittent map.
It is worth observing that all the above-mentioned examples of deterministic
dynamics exhibiting anomalous as well as normal diffusion concern 
chaotic maps, i.e. time-discrete unpredictable dynamics. Another example
of diffusion created by a deterministic dynamics, that of a one-dimensional
Brownian particle subject to elastic collisions with `light' particles
was given in \cite{SzT}. Depending on the asymptotic scaling of the mass ratio
between the test particle and the other particles, the asymptotic process was
shown to be either Ornstein-Uhlenbeck or Wiener.

In this paper we consider the purely mechanical problem of an overdamped
charged particle moving along a line and submitted to an 
electric field and to a random potential created by
a set of quenched random scatterers. 
At variance with similar models that have been recently investigated
\cite{DH,KL}~, we do not include any
random time--dependent force of the Langevin type.

We show that if the electric field is larger than a 
critical value, a current is created and the particle
exhibits, on top of the ballistic motion, a diffusive behavior. 
The diffusion constant can be explicitly
computed. It is a function of the average distance between scatterers
and of the mean value and the variance of the passage time
through a scatterer, and it vanishes, as expected,
with vanishing randomness.
Below the critical value of the electric field also
the current vanishes. The transition from an insulator
to a conductor induced by the electric field
can be described by analogy with phase transitions as being
first or second order. In the former case, the current and
the diffusion constant do not vanish and remain finite at the
threshold, whereas in the latter case the current vanishes
and the diffusion constant diverges according to a power-law
behavior. The critical properties can be characterized by a
scaling law for the power spectrum of the current.
Our study shows that quenched disorder is sufficient to create
normal diffusion and also anomalous diffusion, at least for the
critical value of the electric field.

We note that our model can be interpreted as the
one-dimensional version of the following problem: a massive particle
submitted to a strong friction is sitting or sliding on a rough surface, inclined
with a certain angle $\alpha$ with respect to the ground.
This model problem seems to be important in understanding the phenomenon
of segregation by flow. A variant of it was recently studied numerically and
through a stochastic model \cite{DSB}. In our treatment
the component of the gravity force parallel
to the surface plays the role of the driving field.
The irregularities of the surface, as felt by the particle, are modelled
by a random potential. It is only when $\alpha$ passes through
a certain critical value that the particle will be able to fall down
indefinitely, and then its motion will exhibit a diffusive behaviour,
that turns out to be anomalous at the critical value of the angle.

In the next section we derive the general formulae which will be
used in section 3 for solving two particular models. In the first model,
that we call a gas (or, better, a glass, since frozen in), the scatterers are
distributed uniformly in an interval. In the second model, which 
can be considered as a special case of the first one, and called a crystal,
they follow each other at equal distances. In section 4 we discuss the
transition between conduction and localization through a specific example.
The derivation of a scaling function for the velocity power spectrum 
and some concluding  remarks are contained in section 5.

\section{General setup}

The equation of motion of an overdamped particle in a constant external 
field $E>0$ is
\begin{equation}
\dot x = E - V'(x)
\end{equation}
Throughout the paper we suppose the potential to be a superposition of
disjoint scatterers,
\begin{equation}
V(x)=\sum_{j=1}^N \phi_j(x-r_j) \quad, \quad \phi_j(x) = 0 \quad \quad
\mbox{if} \quad \quad |x|> a
\end{equation}
with centers $r_j$ in an interval $[L_0,L_1]$, subject to
the constraints
\begin{equation}\label{rj}
r_{j+1}-r_j \geq \sigma \geq 2a
\label{cond1}
\end{equation}
This also implies $L_1-L_0\geq (N-1)\sigma$.
We introduce the definitions
\begin{equation}
r_j^{\pm} = r_j \pm a
\end{equation}
and
\begin{equation}
t_j^{\pm} \quad  :  \quad x(t_j^{\pm}) = r_j^{\pm}
\end{equation}
and suppose
\begin{equation}
t_j^- < t_j^+ < t_{j+1}^-
\end{equation}
that is, the particle moves steadily from left to right, `above' the barriers.
This imposes a condition on the external field, namely
\begin{equation}
E>\phi_j'(x)\ \qquad\mbox{for all $j$ and $x$}
\end{equation}
The particle starts at $t=0$ in $x(0)=r_0<r_1^-$ and
for $t < t_1^-$ it moves with a constant speed $\dot x = E$, so that
\begin{equation}
x(t) = r_0 +Et
\end{equation}
By solving for t we obtain,
\begin{equation}
t_1^- = {{r_1^- - r_0}\over{E}}
\end{equation}
If $t_j^- < t < t_j^+$, i.e. $r_j^- < x < r_j^+$,
the equation of motion reads
\begin{equation}\label{xdotx}
{\d\over{\d t}} (x-r_j) = E - \phi_j'(x-r_j)
\end{equation}
whose solution is
\begin{equation}
t - t_j^- = \int_{-a}^{x-r_j} {{\d\eta}\over{E - \phi_j'(\eta)}}
=:\vartheta_j(x-r_j)
\label{tx}
\end{equation}
In particular, the passage time above the $j$-th scatterer is
\begin{equation}\label{tau}
\tau_j:=t_j^+ - t_j^-=\vartheta_j(a)
= \int_{-a}^a {{\d\eta}\over{E - \phi_j'(\eta)}}
\end{equation}
The solution between successive bumps $r_j^+ < x < r_{j+1}^-$,
i.e. $t_j^+ < t < t_{j+1}^-$
again corresponds to the free case $ \dot x = E$, so that
\begin{equation}
t_{j+1}^- - t_j^+ = {{r_{j+1}^- - r_j^+}\over{E}}
\end{equation}
Finally, for $t > t_N^+$ we are again in the free case and
integration yields
\begin{equation}
x = r_N^+ + E(t - t_N^+)
\end{equation}
For the running solutions that we are looking for, the overall equation 
of motion can be put in the form
\begin{equation}
\dot x = E - \sum \phi_j' (x(t)- r_j) \chi_{(t_j^-,t_j^+)}(t)
\end{equation}
where $\chi$ denotes the characteristic function.
The Laplace transform of this equation reads
\begin{equation}
\int_0^{\infty} \dot x(t) e^{-\mu t} \d t =
{E\over\mu}
- \sum_{j=1}^N
\int_{t_j^-}^{t_j^+} e^{-\mu t} \phi_j'(x(t)-r_j) \d t
\end{equation}
Substituting $\phi_j'(x-r_j)$ from (\ref{xdotx}) and using the definitions (\ref{tx}), (\ref{tau}) and
\begin{equation}\label{alpha}
\alpha_j(\mu) =\int_{-a}^a e^{-\mu \vartheta_j(y)}\d y
\end{equation}
we obtain
\begin{equation}
\int_0^\infty \dot x(t) e^{-\mu t} \d t = {E\over\mu} + \sum_{j=1}^N
e^{-\mu t_j^-}\left[\alpha_j - {E\over\mu}\left(1 - e^{-\mu\tau_j}\right)\right]
\end{equation}
or, with
\begin{equation}\label{C}
C_j(\mu)=
\mu \alpha_j(\mu) - E (1 - e^{-\mu \tau_j})
\end{equation}
\begin{equation}
\mu \int_0^{\infty} \dot x(t) e^{-\mu t} \d t
= E + \sum_{j=1}^N e^{-\mu t_j^-}C_j(\mu)
\label{eqmu}
\end{equation}
Time average of the velocity will be obtained by sending first $N$ to 
infinity and then $\mu$ to zero in the last equation.
Since the distance freely run over by the particle up to time 
$t_j^-$ is $r_j-r_0-a-(j-1)2a$, we obtain
\begin{equation}
t_j^- = \sum_{k=1}^{j-1} \tau_k + {{r_j}\over{E}}
- {{(2j -1) a + r_0}\over{E}}
\label{tjm}
\end{equation}
We can take $r_j$ random or not: if they are random the
condition (\ref{cond1}) must be fulfilled. Also $\phi_j$
can be random or not: if they are random  we suppose
they are identically distributed and independent, and also
independent of all $r_i$.
By averaging (\ref{eqmu}) over disorder and noticing that 
$e^{-\mu t_j^-}$ is independent of $C_j(\mu)$,
and $C_j(\mu)$ are identically distributed, we obtain
\begin{equation}\label{aveqmu}
\mu \int_0^{\infty} \overline{\dot x(t)} e^{-\mu t} \d t =
E +
\overline{C(\mu)}
\sum_{j=1}^N
\overline{e^{-\mu t_j^-}}\equiv E+B_N(\mu)
\end{equation}
In what follows, we drop the subscript of averaged quantities whenever 
the average is independent of the subscript.
By using (\ref{tjm}) we can write
\begin{equation}
\overline{e^{-\mu t_j^-}} = e^{(\mu /E)[(2j-1)a +r_0]} D(\mu)^{j-1}
\overline{e^{-\mu r_j/E}}\qquad D(\mu)=
\overline{e^{-\mu \tau}}
\label{emu}
\end{equation}
The average factorizes and becomes a power because $\tau_k$ and 
hence $e^{-\mu \tau_k}$ are independent and identically distributed.

A relevant quantity we are looking for is the time--autocorrelation
function of the velocity. The average of the Laplace transform of this quantity
can be inferred from equations (\ref{eqmu}) and (\ref{aveqmu}) and reads
\begin{eqnarray}\label{mumu'}
\lefteqn{\mu \mu' \int_0^{\infty} \int_0^{\infty} e^{-\mu t - \mu' t'}
[\overline{{\dot x}(t){\dot x}(t')}-\overline{{\dot x}(t)}\,\,
\overline{{\dot x}(t')}] \d t \, \d t'}
\nonumber\\
&=&\sum_{j,j'=1}^N \left[\,\, \overline{e^{-\mu t_j^-} e^{-\mu' t_{j'}^-}
C_j(\mu) C_{j'} (\mu')}
- \overline{e^{-\mu t_j^-}}\,\, \overline{C(\mu)} \overline{e^{-\mu' t_{j'}^-}}\,\,
\overline{C(\mu')}\,\,\right]\nonumber \\
&=&
 \overline{C(\mu) C(\mu')}
\sum_{j=1}^N \overline{e^{-(\mu + \mu') t_j^-}}
+ \Lambda (\mu,\mu') + \Lambda(\mu' , \mu)-B_N(\mu)B_N(\mu')\nonumber\\
\phantom{a}
\end{eqnarray}
Here
\begin{eqnarray}
\lefteqn{
\Lambda(\mu , \mu')= \sum_{1\leq j< j'\leq N}
\overline{ e^{-\mu t_j^- - \mu' t_{j'}^-}
           C_j(\mu) C_{j'}(\mu')              }           }\nonumber\\
&=&
\overline{C(\mu')}\,\,
\overline{e^{-\mu' \tau} C(\mu)}
\sum_{1 \leq j < j' \leq N}
e^{(\mu/E)[(2j-1)a + r_0]}
e^{(\mu'/E) [(2j' -1)a + r_0]}
\nonumber\\
& &\overline{e^{-(\mu r_j+ \mu'r_{j'})/E}}\,\,
D(\mu+\mu')^{j-1}\,\,
D(\mu')^{j'-j-1}\,\,
\end{eqnarray}

\section{Models}
\subsection{A gas of scatterers}

In the first model we are going to consider the scatterers are uniformly 
distributed in the interval $[L_0,L_1]$, but respect the constraints 
(\ref{cond1}).
We introduce a new set of random variables $\{x_j\}$ through
\begin{equation}\label{xj}
r_j=L_0+(j-1)\sigma+x_j
\end{equation}
The joint probability density of $x_1,\ldots,x_N$ is chosen to be
\begin{equation}
p(x_1,\ldots,x_N)=\frac{N!}{L^N}\prod_{j=0}^N\theta(x_{j+1}-x_j)
\end{equation}
where $x_0=0$ and $x_{N+1}=L:=L_1-L_0-(N-1)\sigma$. Then $p$ is 
two-valued and nonzero if and only if
$0\leq x_1\leq x_2\leq\cdots\leq x_N\leq L$.

To compute the average over $x_1,\ldots,x_N$ it is useful to introduce
\begin{equation}
\rho_j(x):=\overline{\delta(x-x_j)}=
{N!\over L^N}{x^{j-1}\over(j-1)!}{(L-x)^{N-j}\over(N-j)!}
\end{equation}
and, for $j<j'$
\begin{equation}\label{rojj'}
\rho_{jj'}(x,x'):=\overline{\delta(x-x_j)\delta(x'-x_{j'})}=
\frac{N!}{L^N}\frac{x^{j-1}}{(j-1)!}
\frac{(x'-x)^{j'-j-1}}{(j'-j-1)!}\frac{(L-x')^{N-j'}}{(N-j')!}
\end{equation}
First we calculate the mean asymptotic velocity. From (\ref{emu}) and
(\ref{xj}) we obtain
\begin{equation}
\sum_{j=1}^N \overline{e^{-\mu t_j^-}}
=e^{-\mu c/E}\int_0^L \d x\, e^{-\mu x/E}\sum_j \rho_j(x)
e^{-\beta(\mu)(j-1)}
\end{equation}
where $c=L_0-a-r_0$ and
\begin{equation}\label{beta}
e^{-\beta(\mu)}\equiv e^{-{\mu\over E}(\sigma-2a)} D(\mu)
\end{equation}
We can choose $r_0=L_0-a$, and thus $c=0$, without restricting generality.
Summation over $j$ can be performed by the use of the binomial formula. It yields
\begin{equation}
\sum_{j=1}^N \overline{e^{-\mu t_j^-}}=
\frac{N}{L}\int_0^L\d x\, e^{-{\mu\over E} x}[1-\frac{x}{L}(1-e^{-\beta})]^{N-1}
\end{equation}
We send $N$ and $L_1-L_0$ to infinity so that the mean distance $\ell$ exists and
$\ell=\lim (L_1-L_0)/N>\sigma$. This yields $\lim N/L=(\ell-\sigma)^{-1}$ and
\begin{equation}\label{expmu}
\sum_{j=1}^\infty \overline{e^{-\mu t_j^-}}=
\left[{\mu\over E}(\ell-\sigma)+1-e^{-\beta(\mu)}\right]^{-1}
\end{equation}
With (\ref{aveqmu}) and the notation
\begin{equation}\label{Bmu}
B(\mu)=\lim B_N(\mu)=\overline{C(\mu)}
\left[{\mu\over E}(\ell-\sigma)+1-e^{-\beta(\mu)}\right]^{-1}
\end{equation}
we can write
\begin{equation}\label{muav}
\mu\int_0^\infty \overline{\dot x(t)}e^{-\mu t}=E+B(\mu)
\end{equation}
When $\mu$ goes to zero we asymptotically find
\begin{eqnarray}\label{asymp}
\overline{C(\mu)}=\mu[2a-E\,\overline{\tau}\,]\nonumber\\
\beta(\mu)=\mu[\,\overline{\tau}+(\sigma-2a)/E]
\end{eqnarray}
and the limit of (\ref{muav}) is therefore
\begin{equation}\label{xinfty}
\overline{\dot x(\infty)}=\frac{E\ell}{\ell+E\,\overline{\tau}-2a}
\end{equation}
This is the time average of the velocity over an infinite run, that is,
\begin{equation}
\overline{\dot x(\infty)}=\lim_{T\to\infty}{1\over T}\int_0^T \dot x(t)\d t
=\lim_{T\to\infty} {x(T)-x(0)\over T}
\end{equation}
Remarkably, this is different from the average of $\dot x(x)$ over 
an infinite distance,
\begin{equation}
\overline{v(x)}\equiv\lim_{r\to\infty}{1\over r}\int_{r_0}^{r_0+r}(E-V'(x))\d x=E
\end{equation}
which is the same as in the absence of the random potential. The reason is
that the work of each scatterer exerted on the particle is vanishing,
$$\phi_j(r_j^+ +0)-\phi_j(r_j^- -0)=0\ . $$
Notice that the result (\ref{xinfty}) could have been obtained without
any computation: $\overline{\dot x(\infty)}$ is the average velocity over
any interval of length $\ell$ containing (the support of) a single scatterer.

To compute the time--autocorrelation function of the velocity we need to evaluate $\Lambda(\mu,\mu')$.
We first rewrite it as
\begin{eqnarray}
\Lambda(\mu , \mu')=
\overline{C(\mu')}\,\,
\overline{e^{-\mu' \tau} C(\mu)}
\sum_{1 \leq j < j' \leq N}
e^{-{\mu\over E}(\sigma -2a) (j-1)}
e^{-{\mu'\over E}(\sigma -2a) (j'-1)}
\nonumber\\
\int_0^L \d x\int_x^L \d x'
\rho_{j j'}(x,x')
e^{-{\mu\over E} x}\,\,e^{-{\mu'\over E} x'}\,\,
D(\mu+\mu')^{j-1}\,\,
D(\mu')^{j'-j-1}\,\,
\end{eqnarray}
With (\ref{beta}),
\begin{eqnarray}
\Lambda(\mu , \mu')=
\overline{C(\mu')}\,\,
\overline{e^{-\mu' \tau} C(\mu)}e^{-{\mu'\over E}(\sigma-2a)}
\int_0^L\d x \int_x^L \d x' e^{-{\mu\over E} x}\,\,e^{-{\mu'\over E} x'}
\nonumber\\
\sum_{1 \leq j < j' \leq N}
\rho_{j j'}(x,x')
e^{-\beta(\mu+\mu')(j-1)}e^{-\beta(\mu')(j'-j-1)}
\end{eqnarray}
Inserting $\rho_{jj'}$ from (\ref{rojj'}) and using the trinomial formula
\begin{equation}
\sum_{0\leq l\leq m\leq M}\frac{M!}{l!(m-l)!(M-m)!}\,\,a^lb^{m-l}c^{M-m}=(a+b+c)^M
\end{equation}
with $l=j-1$, $m=j'-2$, $M=N-2$,
$$a=e^{-\beta(\mu+\mu')}x\ ,\quad
b=e^{-\beta(\mu')}(x'-x)\ ,\quad c=L-x'$$
we get
\begin{eqnarray}
\lefteqn{
\Lambda(\mu , \mu')=
\frac{N(N-1)}{L^2}\,
\overline{C(\mu')}\,\,
\overline{e^{-\mu' \tau} C(\mu)}
e^{-{\mu'\over E} (\sigma -2a)}\times}\\
&&\int_0^L\d x \int_x^L \d x' e^{-{\mu\over E} x}\,\,e^{-{\mu'\over E} x'}
\left\{1+\frac{1}{L}\left[e^{-\beta(\mu+\mu')}x+
e^{-\beta(\mu')}(x'-x)-x'\right]
\right\}^{N-2}\nonumber
\end{eqnarray}
In the limit of $N$ and $L$ going to infinity this yields
\begin{eqnarray}
\lefteqn{\Lambda(\mu , \mu')=
(\ell-\sigma)^{-2}\,\,
\overline{C(\mu')}\,\,
\overline{e^{-\mu' \tau} C(\mu)}
e^{-{\mu'\over E} (\sigma -2a)}=}\\
&&\int_0^\infty\d x \int_x^\infty \d x' e^{-{\mu\over E} x-{\mu'\over E} x'}
\exp\left\{\frac{1}{\ell-\sigma}
\left[
e^{-\beta(\mu+\mu')}x+
e^{-\beta(\mu')}(x'-x)-x'
\right]\right\}\nonumber
\end{eqnarray}
Performing the integral we finally arrive at
\begin{eqnarray}
\Lambda(\mu , \mu')&=&{
\overline{C(\mu')}\,\,
\overline{e^{-\mu' \tau} C(\mu)}\,\,
e^{-{\mu'\over E} (\sigma -2a)}\over
\left[{\mu'\over E}(\ell-\sigma)+1-e^{-\beta(\mu')}\right]
\left[{\mu+\mu'\over E}(\ell-\sigma)+1-e^{-\beta(\mu+\mu')}\right]}\nonumber\\
&=& e^{-{\mu'\over E}(\sigma-2a)}\left[\,\,\overline{e^{-\mu' \tau} C(\mu)}/
\overline{C(\mu+\mu')}\,\,\right]
B(\mu')B(\mu+\mu')
\end{eqnarray}

For two random processes $f(t)$ and $g(t)$ we introduce the notation
\begin{equation}\label{Kfg}
K_{f|g}(\mu,\mu')=\int_0^\infty \int_0^\infty\d t\,\d t' e^{-\mu t-\mu't'}
[\,\,\overline{f(t)g(t')}-\overline{f(t)}\,\,\overline{g(t')}\,\,]
\end{equation}
provided the double integral exists. From equation (\ref{mumu'}) and the subsequent computation
we find in the limit of $N$ going to infinity
\begin{eqnarray}\label{K}
K_{\dot x|\dot x}(\mu,\mu')
&=&{B(\mu+\mu')\over \mu\mu'\,\overline{C(\mu+\mu')}^{\phantom{a}}}
\left[
\,\overline{C(\mu)C(\mu')}+e^{-{\mu'\over E}(\sigma-2a)}\,\,\overline{e^{-\mu'\tau}C(\mu)}
B(\mu')\right.\nonumber\\
&+& \left. e^{-{\mu\over E}(\sigma-2a)}\,\,\overline{e^{-\mu\tau}C(\mu')}
B(\mu)\right]-{B(\mu)B(\mu')\over \mu\mu'}
\end{eqnarray}

Set now $\mu=\epsilon/2+i\omega$ and $\mu'=\mu^*$.
The velocity power spectrum is
\begin{eqnarray}\label{S}
\lefteqn{S_{\dot x|\dot x}(\omega)=\lim_{\epsilon\to 0}\epsilon
K_{\dot x|\dot x}(\mu,\mu^*)=\lim_{\epsilon\to 0}\epsilon\,\,
\overline{\left|\int_0^\infty e^{-\mu t}[\dot x(t)-\overline{\dot x(t)}\,]\d t
\right|^2}
}\nonumber\\
&=& \left(\lim_{\epsilon\to 0}\frac{\epsilon B(\epsilon)}{\overline{C(\epsilon)}}\right)
\frac{1}{\omega^2}
\left\{\overline{|C(i\omega)|^2}+2 {\rm Re}\left[e^{{i\omega\over E}(\sigma-2a)}B(-i\omega)
\overline{e^{i\omega\tau}C(i\omega)}\,\,\right]\right\}\nonumber\\
&=&
{E\over \ell-\overline{A}}\,
\frac{1}{\omega^2}
\left\{\overline{|C(i\omega)|^2}+2 {\rm Re}
\left[\frac{\overline{C(-i\omega)}\,
\overline{e^{i{\omega\over E}(\sigma-A)}C(i\omega)}}
{1-\overline{e^{i{\omega\over E}(\sigma-A)}}-i{\omega\over E}(\ell-\sigma)}
\right]\right\}
\end{eqnarray}
where
\begin{equation}
A = 2a - E\tau \nonumber
\end{equation}
We recall that $C(\mu)$ was defined through equations (\ref{C}), (\ref{alpha}) and (\ref{tx}),
and in the last line of (\ref{S})
the bar stands for averaging over the remaining (single-scatterer) randomness.
Equation (\ref{S}) is the main result of our paper.
For $\omega$ real, $S_{\dot x|\dot x}(\omega)$ is a real, even and nonnegative function.
If $E>E_c=\sup_{j,x}\phi_j'(x)$, the passage times $\tau_j$ are distributed on a bounded support,
the velocity correlations decay exponentially  and
$S_{\dot x|\dot x}(\omega)$ is also
analytic at $\omega=0$. In this case the asymptotic displacement of 
the particle is a drift with a
superimposed normal diffusion. Indeed, the diffusion constant $D$ is given by
\begin{eqnarray}
\lefteqn{2D=\lim_{\epsilon\downarrow 0}\epsilon^2\int_0^\infty  e^{-\epsilon t}
\overline{[x(t)-\overline{x(t)}]^2} \d t} 
\nonumber\\
&=&\lim_{\epsilon\downarrow 0}\int_0^\infty\frac{\epsilon K_{\dot x|\dot x}\left(\frac{\epsilon}{2}(1+iy),
\frac{\epsilon}{2}(1-iy)\right)}{1+y^2}\frac{\d y}{\pi}=S_{\dot x|\dot x}(0)
\end{eqnarray}
Notice that $\epsilon K_{\dot x|\dot x}$ in the integrand has a uniform upper bound.
By using equations
(\ref{aveqmu}), (\ref{emu}), (\ref{beta}), (\ref{expmu}) and (\ref{asymp}), 
a somewhat tedious computation yields
\begin{equation}\label{S0}
2D=S_{\dot x|\dot x}(0)=
\\\frac{E}{(\ell-\overline{A})^3}
\left[
\ell^2(\overline{A^2}-
\overline{A}^2)
+ (\ell-\sigma)^2\,\overline{A}^2
\right]
\end{equation}
Thus, $D$ depends on the randomness through the averages $\ell$, $\overline{\tau}$ and
$\overline{\tau^2}$.
We conclude that in our example of a deterministic dynamics, with a
uniformly distributed
set of quenched random scatterers, there is normal diffusion.
If $E=\sup_{j,x}\phi_j'(x)$, we can loose normal diffusion. We will discuss
this phenomenon later on.

\subsection{A crystal of scatterers}

Here we choose $r_j=L_0+(j-1)\sigma$, that is, the scatterers are placed equi\-distantly.
Because now $r_{j+1}-r_j\equiv\sigma$,
in the limit
when $N$ goes to infinity we also obtain $\ell=\sigma$. This can directly be 
substituted in (\ref{xinfty}), (\ref{S}) and (\ref{S0}) to obtain
\begin{equation}
\overline{\dot x(\infty)}=\frac{E\sigma}{\sigma+E\,\overline{\tau}-2a}
\end{equation}
\begin{equation}\label{S2}
S_{\dot x|\dot x}(\omega)=
{E\over \sigma-\overline{A}}\,
\frac{1}{\omega^2}
\left\{\overline{|C(i\omega)|^2}+2 {\rm Re}
\left[\frac{\overline{C(-i\omega)}\,
\overline{e^{i{\omega\over E}(\sigma-A)}C(i\omega)}}
{1-\overline{e^{i{\omega\over E}(\sigma-A)}}}
\right]\right\}
\end{equation}
and
\begin{equation}
S_{\dot x|\dot x}(0)=
E\sigma^2
\frac{\overline{A^2}-\overline{A}^2}
{(\sigma-\overline{A})^3}
\end{equation}
We note that substitution of $\ell=\sigma$ in equations (\ref{K}) and (\ref{S}) eliminates, at least
asymptotically, the randomness of $r_j$ (still admitting $x_N=o(L)$). If we also drop averaging
over the potentials $\phi_j$, all randomness is lifted, and by the general definition (\ref{Kfg}),
$K_{\dot x|\dot x}$ and $S_{\dot x|\dot x}$ must identically vanish.
It can be easily verified that formulae (\ref{K}) and (\ref{S}) or (\ref{S2}), 
and also (\ref{S0}), indeed show this property.

\section{From conduction to localization}

When $E$ varies continuously and passes the value
$E_c=\sup_{j,x}\phi_j'(x)$, it switches between a conducting state for
$E>E_c$ and an isolating, or localizing, state for $E<E_c$.
At $E=E_c$ there may be localization, if $\overline{\tau}=\infty$
(see equation (\ref{xinfty})~)~, and there may also be conduction
if $\overline{\tau}$ remains finite.
In both cases there may still occur a large variety of different situations,
characterized by different critical exponents and normal or anomalous diffusion.
The transition between conduction and localization bears a resemblence with
phase transitions. For instance, $E$ can be considered as the analog of the
temperature $T$, the region $E>E_c$ that of
$T<T_c$, and $\overline{\dot x(\infty)}$ may correspond, e.g., to the 
spontaneous magnetization and the diffusion constant
to the static zero-field
magnetic susceptibility. We may have first and second-order transitions and
varying exponents depending on the form of probability distribution of the
passage time $\tau$.

Let us discuss
a simple example which illustrates the different possibilities.
We consider
\begin{equation}\label{modelpot}
\phi_j(x)=  f_j \times (|x|-a) \qquad , \qquad |x|\leq a
\end{equation}
so that $\phi_j'(x)=\pm f_j$. Then
\begin{equation}\label{modeltau}
\tau_j
= a \left(\frac{1}{E-f_j} + \frac{1}{E+f_j}\right)
\end{equation}
Suppose that the common probability density 
of the random forces $f_j$ 
has a bounded support $[b,c]$, where $c>0$ and $|b|<c$.
Boundedness is needed to have a transition, and with the
above choice $E_c=c$.
Let us consider on this support a one-parameter family of probability densities
\begin{equation}\label{prob}
p_\gamma(u)=\frac{(\gamma+1)(c-u)^\gamma}{(c-b)^{\gamma+1}}
\end{equation}
with $\gamma>-1$. For $E>c$, 
\begin{equation}
\overline{\tau^n}=\frac{(\gamma+1)a^n}{(c-b)^{\gamma+1}}
\int_0^{c-b}v^\gamma\left[\frac{1}{E+c-v}+\frac{1}{v+\varepsilon}\right]^n\d v
\end{equation}
where we have introduced the short-hand notation $\varepsilon=E-c$.
As $\varepsilon$ goes to zero, asymptotically
\begin{eqnarray}
\overline{\tau^n}&=&
\frac{(\gamma+1)a^n}{(c-b)^{\gamma+1}}
\int_0^{c-b}v^\gamma\left[\frac{1}{2c-v}+\frac{1}{v}\right]^n\d v+o(1)
\quad,\ n<\gamma+1\nonumber\\&=&
\frac{(\gamma+1)a^{\gamma+1}}{(c-b)^{\gamma+1}}
\ln\frac{c-b}{\varepsilon}+O(1)\quad,\ n=\gamma+1\nonumber\\&=&
\frac{(\gamma+1)a^n}{(c-b)^{\gamma+1}}
\frac{\varepsilon^{-n+\gamma+1}}{n-\gamma-1}+
O(\varepsilon^{-n+\gamma+2})\quad,\ n>\gamma+1
\end{eqnarray}
In particular,
\begin{eqnarray}\label{taubar}
\overline{\tau}&=&
\frac{(\gamma+1)a}{(c-b)^{\gamma+1}}
\int_0^{c-b}v^\gamma\left[\frac{1}{2c-v}+\frac{1}{v}\right]\d v+o(1)
\quad,\ \gamma>0\nonumber\\&=&
\frac{a}{c-b}
\ln\frac{c-b}{\varepsilon}+O(1)\quad,\ \gamma=0\nonumber\\&=&
\frac{(1-|\gamma|)a}{|\gamma|(c-b)^{1-|\gamma|}}
\varepsilon^{-|\gamma|}+
O(\varepsilon^{1-|\gamma|})\quad,\ \gamma<0
\end{eqnarray}
and
\begin{eqnarray}
\overline{\tau^2}&=&
\frac{(\gamma+1)a^2}{(c-b)^{\gamma+1}}
\int_0^{c-b}v^\gamma\left[\frac{1}{2c-v}+\frac{1}{v}\right]^2\d v+o(1)
\quad,\ \gamma>1\nonumber\\&=&
\frac{(2a^2}{(c-b)^2}
\ln\frac{c-b}{\varepsilon}+O(1)\quad,\ \gamma=1\nonumber\\&=&
\frac{(\gamma+1)a^2}{(c-b)^{\gamma+1}}
\frac{\varepsilon^{-1+\gamma}}{1-\gamma}+
O(\varepsilon^\gamma)\quad,\ \gamma<1
\end{eqnarray} 
If we are interested in quantities depending on the
random potentials only via $\ell$,
$\overline{\tau}$ and $\overline{\tau^2}$, as $\overline{\dot x(\infty)}$
and $S_{\dot x|\dot x}(0)$, we can distinguish the following cases.

1. $\gamma>1$. In this case $\overline{\tau}$ and $\overline{\tau^2}$ have
a finite limit as $E\downarrow E_c$. As a consequence, 
$\overline{\dot x(\infty)}$ 
is positive and $S_{\dot x|\dot x}(0)$ 
is finite at $E=E_c$ and their value is given respectively by 
(\ref{xinfty}) and (\ref{S0}): there is conduction with normal diffusion.
So when $E$ increases and goes through $E_c$, both
$\overline{\dot x(\infty)}$ and $S_{\dot x|\dot x}(0)$ change discontinuously 
from 0 to a positive value and then vary continuously. This is analogous
with a first order phase transition.

2. $0<\gamma\leq 1$. From the point of view of 
$\overline{\dot x(\infty)}$ the transition is still of first order, but the
divergence of $D$,
\begin{equation}
D\approx\frac{E^3\ell^2\overline{\tau^2}}
{2(\ell-\overline{A})^3}
\end{equation}
with the diverging $\overline{\tau^2}$ when $E\downarrow E_c$ resembles
the divergence of the susceptibility at $T_c$ in second order
magnetic phase transitions. Thus, in this case we have conduction accompanied
with an anomalous diffusion.

3. $-1<\gamma\leq 0$. Both $\overline{\tau}$ and $\overline{\tau^2}$ diverge
when $E\downarrow E_c$, so 
$\overline{\dot x(\infty)}$ tends to zero and 
\begin{equation}
D\approx\frac{\ell^2\overline{\tau^2}}
{2\overline{\tau}\,^3}\propto \varepsilon^{-1-2\gamma}
\end{equation}
with an additional factor $|\ln\varepsilon|^{-3}$ if $\gamma=0$. 
So $D$
diverges if $\gamma>-1/2$, tends to zero if $-1<\gamma<-1/2$ and to a finite
nonzero limit if $\gamma=-1/2$.

Let us emphasise that the probability density $p_\gamma$ is purely continuous
and, thus, the probability that $f_j=E_c$ for a given $j$ is zero. The probability
that $f_j=E_c$ for {\it any} $j$ is still zero. So with probability 1
the particle will never be stopped,
and $\overline{\dot x(\infty)}=0$ means only that,
with probability 1, $x(t)=o(t)$, i.e.
$x(t)$ increases slower than $t$ . It is in this
way that we can understand the different possibilities of diffusion in the third case.

\section{Scaling at criticality}

When $E>E_c$, $S_{\dot x|\dot x}(\omega)$ is a meromorphic function of $\omega$ with no pole in a neighborhood
of the origin. Analyticity at $\omega=0$ will be lost as $E$ attains its critical value $E_c$.
In this section we derive a scaling
law describing the behaviour of $S_{\dot x|\dot x}(\omega;E)$ when $\omega\to 0$ and $E\downarrow E_c$
simultaneously in such way that $z=\varepsilon/\omega\equiv (E-E_c)/\omega$ is kept fixed. More
specifically, we expect that in this limit
\begin{equation}\label{asy}
S_{\dot x|\dot x}(\omega;E)\approx {r(z)\over\omega}
\end{equation}
Below we prove the above form and find the scaling function $r(z)$.
The model that we use for explicit computations is the same as in section 4,
given by (\ref{modelpot}) and (\ref{prob}), although the conclusions 
certainly hold more generally.
First we would like to give an argument why one can expect the asymptotic form (\ref{asy}).
If $\varepsilon>0$, the correlation function
\begin{equation}
\xi(t,t')=\overline{\dot x(t)\dot x(t')}-\overline{\dot x(t)}\,\overline{\dot x(t')}
\end{equation}
decays exponentially with $|t-t'|$, and the correlation time can be approximated with
$\tau_m=2aE/\varepsilon(E+E_c)$, the maximum passage time through a scatterer
(see  equation (\ref{modeltau})~)~. This allows one to write down at least two
different, qualitatively reasonable, approximations to $\xi(t,t')$, namely
\begin{equation}
\xi_1(t,t')=\xi_0\Theta(\tau_m-|t-t'|)\quad\mbox{and}\quad \xi_2(t,t')=\xi_0e^{-|t-t'|/\tau_m}
\end{equation}
with $\Theta(y)$ being the Heaviside function.
Both lead to a form like (\ref{asy}). From $\xi_1$ we obtain
\begin{equation}
S^{(1)}_{\dot x|\dot x}(\omega;E)\approx {2\xi_0\over\omega}\sin\omega\tau_m\approx{2\xi_0\over\omega}
\sin{a\over z}
\end{equation}
while $\xi_2$ yields
\begin{equation}
S^{(2)}_{\dot x|\dot x}(\omega;E)\approx {2\xi_0\tau_m^{-1}\over\tau_m^{-2}+\omega^2}\approx
{2\xi_0\over\omega}{a/z\over 1+(a/z)^2}
\end{equation}

In deriving $r(z)$ we will consider only
negative values of $\gamma$, so that $0<\alpha\equiv -\gamma<1$, choose $b=0$ for the sake of simplicity
and use the notation
\begin{equation}\label{distr}
P(y)=p_{\gamma}(E_c-y)
= p y^{-\alpha}
\end{equation}
cf. equation (\ref{prob}). We note that only
\begin{equation}\label{props}
\sup_y\, y^\alpha P(y)<\infty\quad\mbox{and}\quad \lim_{y\to 0}y^\alpha P(y)=p\ ,
\end{equation}
where $p$ is defined by the normalization,
will be used below, so instead of (\ref{distr}) we can take any $P(y)$ with the
properties (\ref{props}). Notice that
$\int_0^{E_c} {\rm d}y \frac{P(y)}{y} = \infty$.
Accordingly, we obtain
\begin{equation}\label{avet}
\overline{e^{i\omega \tau}}
= \int_0^{E_c} {\rm d}y P(y)\exp{\left[i\omega a
\left(
\frac{1}{\varepsilon + y} + \frac{1}{\varepsilon +2 E_c -y}\right)\right]}
\end{equation}
and
\begin{equation}\label{avetm1}
\lim_{\omega\to 0}
\frac{(\overline{e^{i\omega \tau}} -1)}{\omega^{1-\alpha}}
= p \int_0^{\infty} {\rm d}s\, s^{-\alpha}
\left[ \exp{\left({i \frac{a}{s + z}}\right)} -1 \right] := g(z)
\end{equation}
Furthermore, from the third line of (\ref{taubar}) we have
\begin{equation}\label{avtaupl}
\overline{\tau} =
\tau_0 \varepsilon^{-\alpha}+O(\varepsilon^{1-\alpha})
\end{equation}
as $\varepsilon\to 0$.
In order to compute the current spectrum other terms have
to be computed, like
\begin{equation}\label{term1}
1 - e^{-\beta(i\omega)} = 1 - \overline{
e^{-i {\omega\over E}(\sigma-A)}} \approx
-\omega^{1-\alpha} \overline{g(z)}
\end{equation}
Next, we need the average of
\begin{equation}\label{cimu}
C(\mu) = \mu \int_{-a}^{a} {\rm d}y e^{-\mu \vartheta (y)}
- E(1-e^{-\mu \tau})
\end{equation}
For the potential (\ref{modelpot})
\begin{equation}\label{teta}
\vartheta (y) =
\int_{-a}^{y} \frac{{\rm d}s}{E - f} = \Theta(-y)
\frac{(a + y)} {E+f} + \Theta(y)
\frac{a}{E+f} +
\frac{y}{E-f}
\end{equation}
Inserting (\ref{teta}) in (\ref{cimu}),
\begin{equation}\label{cimun}
C(\mu) = f \left[ 1 - 2 \exp{\left(-\frac{\mu a}{E + f}\right)}
+ \exp{\left(-\mu a \left(\frac{1}{E+f} + \frac{1}{E-f}\right)\right)}\right]
\end{equation}
so that
\begin{eqnarray}\label{cimuav}
\overline{C(\mu)} &=& \int_0^{E_c} {\rm d}y\, P(y)\,
(E_c - y)
\left[ 1 - 2 \exp{\left(-\frac{\mu a}{\varepsilon + 2 E_c- y}\right)}
\right. \nonumber \\
&+&\left. \exp{\left(-\frac{\mu a}{\varepsilon + y}
-\frac{\mu a}{\varepsilon + 2E_c-y}\right)}\right]
\end{eqnarray}
and 
\begin{eqnarray}\label{ecimuav}
\overline{e^{\mu \tau}C(\mu)} &=& \int_0^{E_c} {\rm d}y\, P(y)\,
(E_c - y)
\left[
\exp{\left(\frac{\mu a}{\varepsilon + y}
+\frac{\mu a}{\varepsilon + 2E_c-y}\right)}
\right.
\nonumber \\
&&\left.
- 2 \exp{\left(\frac{\mu a}{\varepsilon - y}\right)} + 1
\right]
\end{eqnarray}
Accordingly, in the limit of vanishing $\omega$
\begin{equation}
\omega^{\alpha -1} \overline{C(i \omega)} \to p\, E_c\,
\int_0^{\infty} {\rm d}s\, s^{-\alpha}\, \left[
\exp{\left(-i \frac{a}{s + z}\right)} -1 \right] = E_c \, g^*(z)
\end{equation}
and
\begin{equation}
\omega^{\alpha -1} \overline{e^{i\omega\tau}\,C(i \omega)} \to p\, E_c\,
\int_0^{\infty} {\rm d}s\, s^{-\alpha}\, \left[
\exp{\left(i \frac{a}{s + z}\right)} -1 \right] = E_c \, g(z)
\end{equation}
Moreover, one has (taking $\ell=\sigma$ for simplicity)
\begin{eqnarray}
B(-i \omega) &=& \frac{\overline{C(-i\omega)}}
{1-\exp{(-\beta(i\omega))}} \nonumber \\
&=& \left[-\frac{\omega^{\alpha -1} \overline{C(i \omega)}}
{\omega^{\alpha -1} (1-\exp{(-\beta(i\omega))}}\right]^*
\to \left[- \frac{E_c\, g^*(z)}{g^*(z)}\right]^*
= -E_c
\end{eqnarray}
Similar expressions have to be obtained for
\begin{eqnarray}\label{cimuav2}
\overline{|C(i\omega)|^2} &=& \int_0^{E_c} {\rm d}y\, P(y)\,
(E_c - y)^2
\left|1 - 2 \exp{\left(-i\frac{\omega a}{\varepsilon + 2 E_c- y}\right)}
\right. \nonumber \\
&+&\left. \exp{\left(-i\frac{\omega a}{\varepsilon + y}
-i\frac{\omega a}{\varepsilon + 2E_c-y})\right)}\right|^2
\end{eqnarray}
so that
\begin{equation}\label{ocimuav2}
\omega^{\alpha -1}\overline{|C(i\omega)|^2} \to
p\,E_c^2\int_0^{\infty} {\rm d}s\, s^{-\alpha}\,
\left|\exp{\left(-i\frac{a}{s + z}\right)}
-1 \right|^2 = E_c^2 u(z)
\end{equation}
By substituting into eq. (\ref{S}) we finally obtain
\begin{eqnarray}\label{SEc}
\lefteqn{S_{\dot x|\dot x}(\omega)= }\nonumber \\
&&\frac{\varepsilon + E_c}{\omega^2[\ell-2a+(\varepsilon + E_c)\overline{\tau}]}
\left\{\overline{|C(i\omega)|^2}+2 {\rm Re}\left[e^{i\omega\delta}
\, B(-i\omega)
\overline{e^{i\omega\tau}C(i\omega)}\,\,\right]\right\} \nonumber \\
&\approx&
\frac{\varepsilon + E_c}{\omega^2\omega^{\alpha -1}[\ell-2a+(\varepsilon + E_c)
\tau_0 \varepsilon^{-\alpha}]}
E_c^2 \left[ u(z) - 2 {\rm Re} g(z)\right]
\end{eqnarray}
Accordingly, in the limit of small $\varepsilon$ one obtains
\begin{eqnarray}\label{SEcsm}
\omega^{\alpha +1} \varepsilon^{-\alpha}
S_{\dot x|\dot x}(\omega) &\to&
\frac{E_c^2}{\tau_0} \left[ u(z) - 2 {\rm Re} g(z)\right] \nonumber \\
&=&
\frac{E_c^2}{\tau_0} p \int_0^{\infty}{\rm d}t\, t^{-\alpha}
\left[
\left| 1 - e^{{-i\frac{a}{t+z}}}\right|^2
- 2 {\rm Re} \left(
e^{{i\frac{a}{t+z}}} -1
\right)
\right] \nonumber \\
&=&
\frac{4\, E_c^2}{\tau_0} p \int_0^{\infty}{\rm d}t\, t^{-\alpha}
\left[ 1 - \cos{\left(\frac{a}{t+z}\right)}
\right]
\end{eqnarray}
Finally,
\begin{equation}
\lim_{\omega\to 0} \omega
S_{\dot x|\dot x}(\omega;E_c + \omega z)=
\frac{
4\, E_c^2\, z^{\alpha}\, \int_0^{\infty}{\rm d}t\, t^{-\alpha}
\left[ 1 - \cos{\left(\frac{a}{t+z}\right)} \right]
}
{a\,\int_0^{\infty}
{\rm d}t\, t^{-\alpha} (1+t)^{-1}
 } \equiv r(z)
\end{equation}
It is worth pointing out the relevance of this result: a $\omega^{-1}$
component emerges naturally at criticality in the spectral properties
of a purely mechanical disordered model, that can be viewed as the
classical counterpart of the Anderson's model for localization.

{\bf Acknowledgments}

The authors thank the hospitality of EPF Lausanne (R. L. and A. S.) and of
RISSPO Budapest (H. K. and R. L.) where this work was done. Partial
support from the Hungarian Scientific Research Fund grant T 30543 and
the Center of Excellence ICA1-CT-2000-70029 is gratefully acknowledged.


\begin{thebibliography}{999}
\bibitem{Sinai} Ya. G. Sinai, Theor. Prob. Appl. {\bf 27}, 247 (1982).
\bibitem{BG} Bouchaud J P and Georges A, Phys. Rep.,
{\bf 195}, 127 (1990)
\bibitem{Ott} E. Ott, "Chaos in Dynamical Systems", Cambridge University
Press, Cambridge (1993).
\bibitem{Lichten} A. J. Lichtenberg and M. A. Liebermann, Physica D {\bf 33},
211 (1988).
\bibitem{Ishizaki} R. Ishizaki, T. Horita, T. Kobayashi and H. Mori, Progr. 
Theor. Phys. {\bf 85}, 1013 (1991).
\bibitem{Geisel} T. Geisel, J. Nierwetberg and A. Zacherl, Phys. Rev. Lett.
{\bf 54}, 616 (1985).
\bibitem{Pikov} A. S. Pikovsky, Phys. Rev. A {\bf 43}, 3146 (1991).
\bibitem{SzT} D. Sz\'asz and B. T\'oth, Commun. Math. Phys. {\bf 104}, 445 (1986),
{\it ibid.} {\bf 111}, 41 (1987), J. Stat. Phys. {\bf 47}, 681 (1987).
\bibitem{DH} S.I. Denisov and W. Horsthemke, Phys. Rev. E {\bf 62}, 3311 (2000).
\bibitem{KL} G. Kaniadakis and G. Lapenta, Phys. Rev. E {\bf 62}, 3246 (2000).
\bibitem{DSB} S. Dippel, L. Samson and G. G. Batrouni, in: Traffic and granular
flow: Workshop in J\"ulich 1995, ed. D. E. Wolff et al, World Scientific, 1996
\end{thebibliography}
\end{document}